\documentclass[manuscript]{aastex}

\shorttitle{FIR [O {\sc iii}] lines from $z>8$ galaxies}
\shortauthors{Inoue et al.}

\begin{document}


\title{ALMA will determine the spectroscopic redshift $z>8$ with 
FIR [O {\sc iii}] emission lines} 


\author{A. K. Inoue\altaffilmark{1}, I. Shimizu\altaffilmark{1,2}, 
Y. Tamura\altaffilmark{3}, H. Matsuo\altaffilmark{4},
T. Okamoto\altaffilmark{5}, and N. Yoshida\altaffilmark{6,7}}
\email{akinoue@las.osaka-sandai.ac.jp}


\altaffiltext{1}{College of General Education, Osaka Sangyo University,
3-1-1 Nakagaito, Daito, Osaka 574-8530, Japan}
\altaffiltext{2}{Department of Astronomy, The University of Tokyo, 7-3-1
Hongo, Tokyo 113-0033, Japan}
\altaffiltext{3}{Institute of Astronomy, The University of Tokyo,
Mitaka, Tokyo 181-0015, Japan} 
\altaffiltext{4}{National Astronomical Observatory of Japan, 2-21-1
Osawa, Mitaka, Tokyo 181-8588, Japan}
\altaffiltext{5}{Department of Cosmosciences, Graduate School of
Science, Hokkaido University, N10 W8, Kitaku, Sapporo 060-0810, Japan}
\altaffiltext{6}{Department of Physics, The University of Tokyo, 7-3-1
Hongo, Tokyo 113-0033, Japan}
\altaffiltext{7}{Kavli Institute for the Physics and Mathematics of the
Universe, TODIAS, The University of Tokyo, 5-1-5 Kashiwanoha, Kashiwa,
Chiba 277-8583, Japan}


\begin{abstract}
We investigate the potential use of nebular emission lines in the
 rest-frame far-infrared (FIR) for determining spectroscopic redshift of
 $z>8$ galaxies with the Atacama Large Millimeter/sub-millimeter Array
 (ALMA). 
After making a line emissivity model as a function of metallicity,
 especially for the [O {\sc iii}] 88 $\mu$m line which is likely to be
 the strongest FIR line from H {\sc ii} regions, we predict the line
 fluxes from high-$z$ galaxies based on a cosmological hydrodynamics
 simulation of galaxy formation. 
Since the metallicity of galaxies reaches at $\sim0.2~Z_\odot$ even at
 $z>8$ in our simulation, we expect the [O {\sc iii}] 88 $\mu$m line as
 strong as 1.3 mJy for 27 AB objects, which is detectable at a high
 significance by $<1$ hour integration with ALMA. 
Therefore, the [O {\sc iii}] 88 $\mu$m line would be the best tool to
 confirm the spectroscopic redshifts beyond $z=8$.
\end{abstract}





\section{Introduction}

Finding the highest redshift objects is placing constraints on the
theory of baryonic physics to form luminous objects in the Universe. 
The latest survey with the Hubble Space Telescope (HST) has provided a
number of candidates of $z>7$ galaxies by the so-called drop-out 
technique \citep[e.g.,][]{Ellis2013}. However, the redshifts of these
Lyman break galaxies (LBGs) are not yet confirmed through spectroscopy. 
Since the rest-frame ultraviolet (UV) continuum of the LBGs are too
faint ($>27$ AB) to be detected with current spectrograph, it was often
assumed that Ly$\alpha$ emission is the best tool to confirm the
redshift. Yet, attempts to detect Ly$\alpha$ for $z>8$ LBGs failed 
\citep{Brammer2013,Bunker2013,Capak2013,Treu2013}, indicating that
Ly$\alpha$ was substantially weakened by the intergalactic neutral
hydrogen before the completion of the cosmic reionization. Otherwise the
LBGs were just interlopers \citep{Pirzkal2013,Brammer2013}. If
Ly$\alpha$ emission at $z>8$ is so weakened that we cannot detect, we
should consider other emission lines to confirm their redshift.

Considering the superb ability of the Atacama Large
Millimeter/sub-millimeter Array (ALMA), the rest-frame far-infrared (FIR)
emission lines may be attractive. For example, the [C {\sc ii}] 158
$\mu$m line is very luminous and often detected from high-$z$ 
sources including $z>6$ QSOs \citep{Maiolino2005}. However, the line is
not detected from $z>6$ Ly$\alpha$ emitters
\citep{Walter2012,Kaneker2013,Ouchi2013}, suggesting a different
situation of the interstellar medium in these high-$z$ low-metallicity
galaxies. In addition, the [C {\sc ii}] line at $8.0<z<10.6$, where the
highest-$z$ LBGs reside, is redshifted into the ALMA band 5 not
available soon.

The FIR [O {\sc iii}] lines at 52 and 88 $\mu$m are known as prominent
lines from H {\sc ii} regions since 1970s \citep{Ward1975}, while the
lines have been rarely discussed in the high-$z$ context so far
because of the lack of suitable instruments. The first FIR [O {\sc iii}]
detection from cosmologically distant sources is reported by
\cite{Ferkinhoff2010} from two $z\simeq3$ and 4 gravitationally lensed
dusty AGN/starburst galaxies. In the local Universe, the Infrared Space
Observatory (ISO) and the Japanese Infrared Satellite AKARI detected the
lines from Galactic H {\sc ii} regions \citep{Mizutani2002,Matsuo2009},
from a giant H {\sc ii} region, 30 Doradus, in the Large Magellanic
Cloud (LMC) \citep{Kawada2011}, and from many nearby galaxies
\citep{Brauher2008}. Interestingly, recent {\it Herschel} observations
have revealed that the [O {\sc iii}] 88 $\mu$m line is often stronger
than the [C {\sc ii}] line in low-metallicity nearby dwarf galaxies 
(Figure~5 of Madden et al.~2012; see also Cormier et al.~2012), 
suggesting the usefulness of the [O {\sc iii}] line at high-$z$ where
most galaxies are low-metallicity. Last but not least, the [O {\sc iii}]
88 $\mu$m line at $8.1<z<11.3$ falls into the ALMA band 7 in operation.

In section 2, we construct a FIR nebular emission model in high-$z$
Universe based on our cosmological simulation and the photoionization
code {\sc cloudy}, followed by the expected FIR line fluxes presented
in section 3. Finally, we discuss the feasibility for detecting the
lines from $z>8$ galaxies with ALMA in section 4.

Throughout this Letter, we adopt a $\Lambda$CDM cosmology 
with the matter density $\Omega_{\rm{M}} = 0.27$, 
the cosmological constant $\Omega_{\Lambda} = 0.73$, 
the Hubble constant $h = 0.7$ in the unit of 
$H_0 = 100 {\rm ~km ~s^{-1}~Mpc^{-1}}$ and  
the baryon density $\Omega_{\rm B} = 0.046$. 
The matter density fluctuations are normalized by setting
$\sigma_8 = 0.81$ \citep{WMAP}. 
All magnitudes are quoted in the AB system \citep{Oke1990}.

\section{Model of FIR H II region lines in high-$z$}

In this Letter, we consider only lines from H {\sc ii} regions, and
then, we assume the line luminosity, $L_{\rm line}$, to be proportional
to the instantaneous star formation rate (SFR) of a galaxy, $\dot{M_*}$;
\begin{equation}
 L_{\rm line} = C_{\rm line}(Z,U,n_{\rm H}) \dot{M_*}\,,
\end{equation}
where $C_{\rm line}$ is the line emissivity per unit SFR and depends on
the metallicity, $Z$, the ionization parameter, $U$, and the hydrogen
number density, $n_{\rm H}$, in H {\sc ii} regions of the galaxy. This
assumption is usual for hydrogen recombination lines because the
Lyman continuum (LyC) making H {\sc ii} regions are emitted only by
massive stars whose life-time ($\sim1$ Myr) is short enough to
represent the instantaneous SFR of galaxies
\citep[e.g.,][]{Kennicutt1998}. Given that the same LyC also ionizes the
metal atoms, the similar assumption for metal forbidden lines as
equation (1) would be reasonable. However, the proportional factor,
$C_{\rm line}$, depends on the nebular parameters of $U$ and $n_{\rm H}$
as well as the metallicity $Z$ for the case of forbidden lines. On the
other hand, we avoid modelling photodissociation regions and molecular
clouds surrounding H {\sc ii} regions because it requires more
complex physical and chemical processes
\citep[e.g.,][]{Abel2005,Nagao2011} and may cause a large uncertainty,
although it enables us to predict some strong FIR lines such as [C {\sc
ii}] 158 $\mu$m and [O {\sc i}] 63 $\mu$m. It would be an interesting
future work.

Using the photoionization code {\sc cloudy} version c13.01
\citep{Ferland2013}, we make a model of $C_{\rm line}$. The {\sc cloudy}
calculations are similar to \cite{Inoue2011}. We assume six
metallicities as $\log_{10}(Z/Z_\odot)=-5.3$, $-3.3$, $-1.7$, $-0.7$,
$-0.4$, and 0.0, three ionization parameters as $\log_{10}U=-3.0$,
$-2.0$, and $-1.0$, and four hydrogen number densities as 
$\log_{10}(n_{\rm H}/{\rm cm}^{-3})=0.0$, 1.0, 2.0, and 3.0. 
The ``H {\sc ii} region'' set of the gas elemental abundance based
on the observations of the Orion nebula is used and the ``Orion type''
dust grains are included. The nebular and stellar metallicities
are assumed to be equal. The shape of the stellar spectra are taken from
{\sc Starburst99} \citep{Leitherer1999} and \cite{Schaerer2002},
depending on the metallicity. The initial mass function (IMF) is assumed
to be a Salpeter one with 0.1--100 $M_\odot$. We consider the scenario
of a constant star formation of 10 Myr but the 
age of the star formation has negligible impact on $C_{\rm line}$. 
We also note that negligible differences are found when we adopt another
population synthesis code {\sc p\'{e}gase} ver.~2 \citep{PEGASE}. 
We also assume a constant density throughout the H {\sc ii} regions and
the plain-parallel geometry. The calculations are stopped if the
electron temperature becomes less than $10^{3.5}$ K or the electron
fraction becomes less than 1\%. No escape of the LyC from H {\sc ii}
regions (and galaxies) is considered. The line emissivities are reduced
approximately by a factor of $1-f_{\rm esc}$ when the escape fraction is
$f_{\rm esc}$.

\begin{figure}
 \epsscale{0.7}
 \plotone{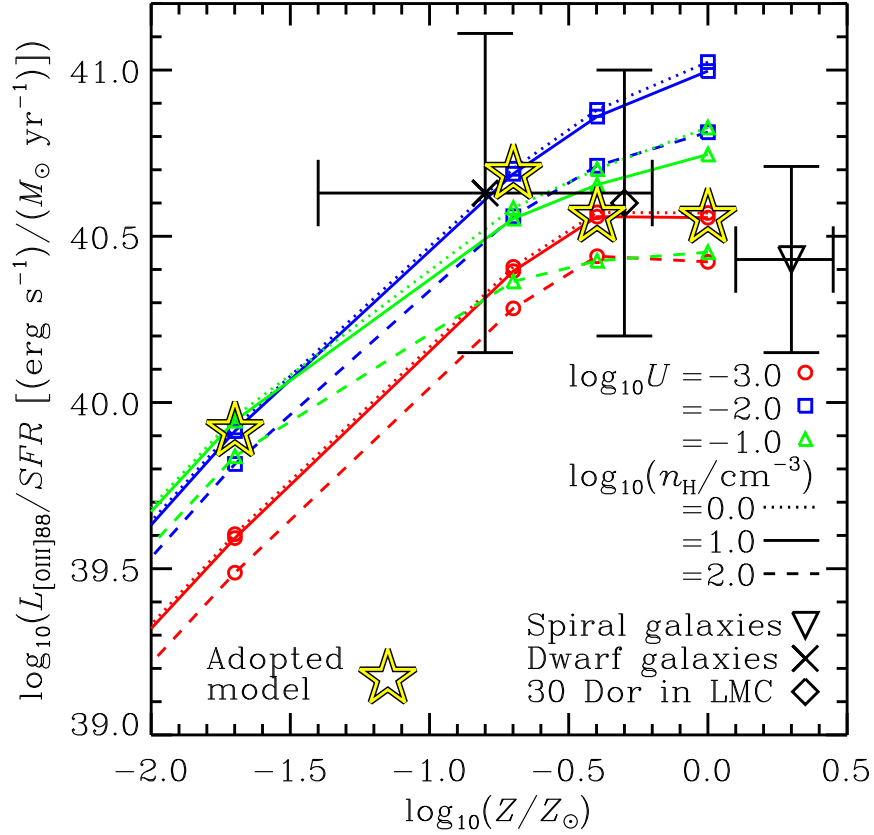}
 \caption{Luminosities of the [O {\sc iii}] 88 $\mu$m line per unit star
 formation rate (a Salpeter initial mass function with 0.1--100
 $M_\odot$) as a function of metallicity for 9 combinations of the
 ionization parameter $\log_{10}U$ and the hydrogen density
 $\log_{10}n_{\rm H}$ indicated in the panel. No escape of the Lyman
 continuum from H {\sc ii} regions is taken into account. The inverse
 triangle, cross, and diamond with error-bars are observational
 estimates of nearby spiral galaxies \citep{Brauher2008}, nearby dwarf
 galaxies \citep{Madden2012,Madden2013}, and 30 Doradus in the Large
 Magellanic Cloud \citep{Kawada2011}, respectively. The large
 five-pointed stars indicate the models adopted in this Letter and
 summarized in Table 1.}
\end{figure}


\begin{deluxetable}{lcccccccccccc}
 \tabletypesize{\scriptsize}
 \tablecaption{Line emissivities per unit star formation rate 
 $C_{\rm line}$ in equation (1) adopted in this Letter. 
 The tabulated values are 
 $\log_{10}(C_{\rm line} [{\rm erg}~{\rm s}^{-1}/(M_\odot~{\rm yr}^{-1})])$. 
 Density is always assumed to be $\log_{10}(n_{\rm H}/{\rm cm}^{-3})=1.0$.}
 \tablehead{\colhead{} & \colhead{} & \colhead{[S {\sc iv}]} 
 & \colhead{[Ne {\sc ii}]} & \colhead{[Ne {\sc iii}]} 
 & \colhead{[S {\sc iii}]} & \colhead{[S {\sc iii}]}
 & \colhead{[Ne {\sc iii}]} & \colhead{[O {\sc iii}]}
 & \colhead{[N {\sc iii}]} & \colhead{[O {\sc iii}]}
 & \colhead{[N {\sc ii}]} & \colhead{[N {\sc ii}]} \\
 \colhead{$Z/Z_\odot$} & \colhead{$\log_{10}U$} & \colhead{10.51} 
 & \colhead{12.81} & \colhead{15.55} & \colhead{18.67} & \colhead{33.47}
 & \colhead{36.01} & \colhead{51.80} & \colhead{57.21} & \colhead{88.33}
 & \colhead{121.7} & \colhead{205.4}}
 \startdata
   1.0 & $-3.0$ & 38.87 & 40.01 & 40.00 & 40.32 & 40.72 & 38.95 
   & 40.32 & 40.14 & 40.56 & 39.81 & 39.82 \\
   0.4 & $-3.0$ & 39.09 & 39.50 & 40.07 & 40.10 & 40.46 & 39.02 
   & 40.33 & 39.70 & 40.56 & 39.04 & 39.07 \\
   0.2 & $-2.0$ & 40.12 & 38.38 & 40.02 & 39.86 & 40.20 & 38.97 
   & 40.46 & 39.52 & 40.69 & 37.67 & 37.72 \\
   0.02 & $-2.0$ & 39.39 & 37.43 & 39.25 & 39.05 & 39.36 & 38.20 
   & 39.69 & 38.74 & 39.91 & 36.79 & 36.85 \\
   $5\times10^{-4}$ & $-2.0$ & 38.07 & 35.58 & 37.75 & 37.42 & 37.73 
   & 36.71 & 38.18 & 37.14 & 38.41 & 35.25 & 35.31 \\
   $5\times10^{-6}$ & $-2.0$ & 36.22 & 33.60 & 35.85 & 35.47 & 35.78
   & 34.81 & 36.29 & 35.19 & 36.52 & 33.36 & 33.42 \\
 \enddata
\end{deluxetable}

From the {\sc cloudy} calculations, we find that the [O {\sc iii}] 88
$\mu$m line is the strongest line for the nebular parameters examined in
this Letter. Figure~1 shows the [O {\sc iii}] line luminosities per unit
SFR as a function of metallicity for 9 combinations of nebular
parameters as indicated in the panel. The luminosities with 
$\log_{10}(n_{\rm H}/{\rm cm}^{-3})=3.0$ are much smaller than those
shown in Figure~1 as a trend found in the figure: a smaller emissivity
for a higher density. The line emissivities for a constant 
ionization parameter are roughly proportional to the metallicity when
$Z<0.1Z_\odot$ as the oxygen abundance increases. When $Z>0.1Z_\odot$,
the dependence becomes weaker because of lower LyC emissivity for higher
metallicity.

For a comparison, we plot observational estimates in Figure~1. 
\cite{Kawada2011} reported the flux ratio of 
$F_{\rm [OIII]88}/F_{\rm H\alpha}\simeq0.5$ to 1.0 in 30 Doradus of the
LMC. Adopting the conversion formula from H$\alpha$ luminosity to SFR by
\cite{Hirashita2003}, we obtain the emissivity shown by the
diamond.\footnote{In the formula of \cite{Hirashita2003}, there are two
parameters: the dust correction factor for H$\alpha$ ($A_{\rm H\alpha}$)
and the hydrogen ionizing fraction in the LyC ($f$). Note that a part of
the Lyman continuum is absorbed by dust before using hydrogen ionization
\citep{Inoue2001a}. We take $A_{\rm H\alpha}=0.5$ and $f=0.7$ for the
LMC from \cite{Inoue2001b}.} The vertical error-bar indicates the
uncertainty of the conversion and the sample variance of the flux
ratio. The metallicity is taken from \cite{vdBergh2000}.
\cite{Madden2012,Madden2013} reported 
$\log_{10}(F_{\rm [OIII]88}/F_{\rm FIR})=-2.04\pm0.29$ for nearby
low-metallicity dwarf galaxies. Converting the FIR luminosity to SFR by
using the formula of \cite{Hirashita2003}, we obtain the cross point
with error-bars.\footnote{First, we converted the FIR (40--120 $\mu$m)
luminosity estimated from IRAS measurements to the total IR (8--1000
$\mu$m) luminosity, assuming the dust temperature of 30 K and the IR
emissivity index of 1.0. Then, the IR luminosity is converted to the SFR
by the formula of \cite{Hirashita2003} \citep[see also][]{Inoue2000}
which has three parameters. We adopt $f=0.8\pm0.2$, $\epsilon=0.3\pm0.2$
and $\eta=0.1\pm0.1$ for low-metallicity starbursting galaxies.} The
vertical error-bar includes the uncertainty of the conversion and the
observed sample variance. The horizontal error-bar indicates the sample
metallicity distribution taken from \cite{Madden2013}. 
\cite{Brauher2008} presented a compilation of [O {\sc iii}] 88 $\mu$m
observations of nearby galaxies. If we select only spiral galaxies
from their sample, we obtain 
$\log_{10}(F_{\rm [OIII]88}/F_{\rm FIR})=-2.61\pm0.20$. Again 
adopting the formula by \cite{Hirashita2003}, we obtain the inverse
triangle with error-bars.\footnote{The FIR to IR conversion is done with
the dust temperature of 30 K and the emissivity index of 1.0. The IR to
SFR conversion is done with the recommended factor for nearby
star-forming galaxies in \cite{Hirashita2003}.} The vertical error-bar
includes the uncertainty of the conversion and the sample variance. The 
metallicity is estimated from the mean absolute magnitude of the sample
galaxies via the correlation between the magnitude and the metallicity
presented by \cite{Tremonti2004}. 

While the uncertainties are still large, we may find a trend that
the [O {\sc iii}] emissivity is slowly decreases as the metallicity
increases from $\approx0.2Z_\odot$. We also find that no single
combination of the nebular parameters of $\log_{10}U$ and
$\log_{10}n_{\rm H}$ reproduces this trend. Then, we consider a model
with a constant $\log_{10}n_{\rm H}$ but a higher $\log_{10}U$ at lower
metallicities. Such a trend may be realized by a higher LyC production
rate and a harder spectrum of lower metallicity stars. We therefore
adopt the models indicated by the large five-pointed stars in this
Letter. However, we should note that this may not be a unique
combination of the parameters compatible with the observations. Table~1 
is a summary of $C_{\rm line}$ for 11 H {\sc ii} region lines calculated
in the adopted models.

In order to predict the FIR line fluxes from high-$z$ galaxies, we need
their SFRs in equation (1). We adopt a cosmological simulation by
\citet{Shimizu2013} which was developed to examine physical properties 
of LBGs at $z\sim7$--10. The simulation code is based on a Tree-PM
smoothed particle hydrodynamics code {\sc gadget-3} updated from 
{\sc gadget-2} \citep{Gadget}. We have implemented star formation,
supernova (SN) feedback and chemical enrichment following
\citet{Okamoto2008, Okamoto2009, Okamoto2010}. We employ 
$N = 2 \times 640^3$ particles for dark matter and gas in a comoving
volume of $50 h^{-1}{\rm ~Mpc}$ cube. The mass of a dark matter 
particle is $3.01 \times 10^7 h^{-1}~M_{\odot}$ and that of a gas
particle is initially $6.09 \times 10^6 h^{-1}~M_{\odot}$. Gas particles
can spawn star particles when they satisfies a set of criteria for star
formation. In each snapshot of the simulation, we run the {\sc subfind}
algorithm \citep{Springel2001} to identify groups of dark matter,
gas, and star particles as galaxies. Parameters in the code such as SN
feedback and dust attenuation are calibrated so as to reproduce the
stellar mass functions and UV luminosity functions observed at
$z\geq7$. We have constructed a light-cone output from a number of
snapshots of the simulation, calculated the apparent magnitudes in a
number of broadband filters, and then, applied the exactly same color
selection criteria as real observations to select LBGs at $z\sim7$, 8,
9, and 10. See \citet{Shimizu2013} for more details.


The FIR line fluxes, $F_{\rm line}$s, for individual galaxies extracted
from the cosmological simulation are estimated by the following
procedure; First, we assign $C_{\rm line}(Z_{\rm neb})$s to each
simulated LBG, where $Z_{\rm neb}$ is the ``nebular''
metallicity\footnote{A weighted mean metallicity of star particles
composing of a galaxy with the LyC luminosity of the particles as the
weight \citep{Shimizu2013}.} of the LBG, by interpolating the values in
Table~1. Then, we obtain the line luminosities by equation~(1) with the
SFR of the simulated LBG. Finally, the luminosities are converted to the
fluxes by the luminosity distance in the light-cone. In addition, the
peak intensities of the lines, $F_{\nu_0}^{\rm peak}$, are calculated by 
the following formula: 
$F_{\nu_0}^{\rm peak}=F_{\rm line}(1+z)c/\nu_0/v_{\rm 1D}/\sqrt{\pi}$, 
where $c$ is the light speed, $z$ is the redshift,$\nu_0$ is the
rest-frame line center frequency, and $v_{\rm 1D}$ is the standard
deviation of the one-dimensional gas velocity which is assumed to be
equal to that of the dark matter.

\section{Expected FIR line fluxes of high-$z$ galaxies}

Figure~2 shows the expected flux of the [O {\sc iii}] 88 $\mu$m line,
which is the strongest among the lines examined in this Letter, from the
simulated LBGs at $z\sim7$ to 10 as a function of the apparent
magnitudes. We find that a good correlation between the line flux and
the apparent rest-frame UV magnitude and it does not change along the
redshift very much. The dispersion is larger for fainter galaxies
because the dispersions of metallicity, SFR, and dust attenuation are
also larger for fainter galaxies in our simulation since the SN feedback
affects them largely and their star formation histories fluctuate more
\citep{Shimizu2013}. 

\begin{figure}
 \epsscale{0.7}
 \plotone{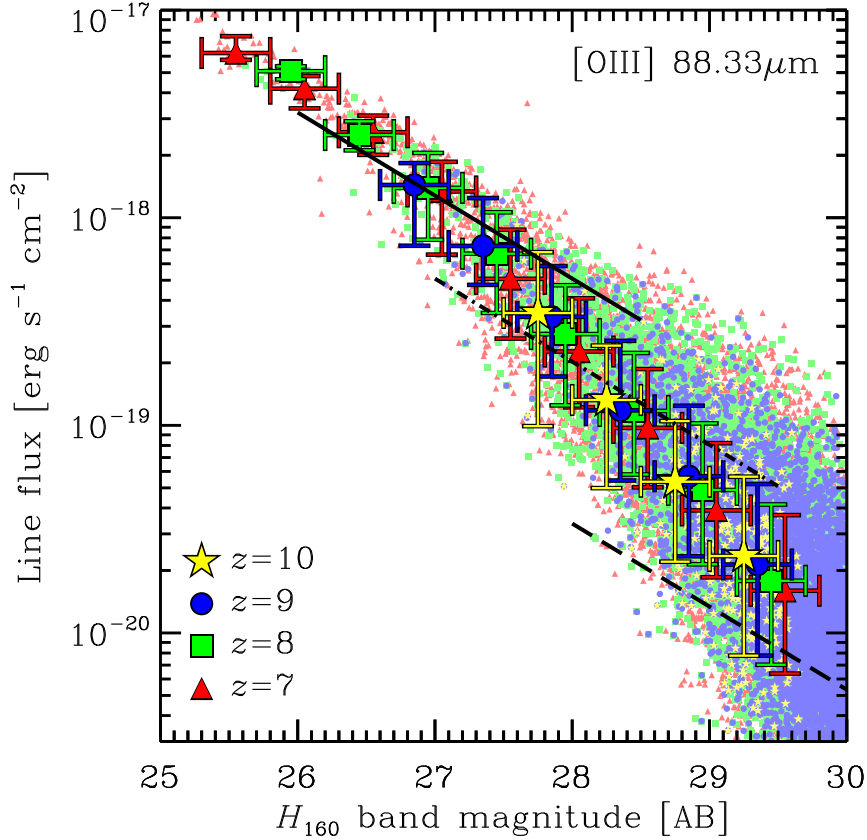}
 \caption{The expected line flux of the [O {\sc iii}] 88 $\mu$m
 line from Lyman break galaxies at $z\sim7$ (triangles), 8 (squares), 9
 (circles), and 10 (stars) in the simulation of \citet{Shimizu2013}. The
 horizontal axis is the simulated apparent HST/WFC3 $H_{160}$ magnitude.
 The large symbols with error-bars show the median values and the
 central 68\% ranges for the galaxies divided into 0.5 magnitude step
 bins shown by the horizontal error-bars. The solid, dot-dashed, and
 dashed lines show the equation (2) for $z=7$, $f_{\rm esc}=0$, and a
 set of $(Z/Z_\odot,A_{\rm UV})=(0.2,1.0)$, (0.2,0.0), and (0.02,0.0), 
 respectively.}
\end{figure}

Relating the SFR in equation~(1) to the UV magnitude, we can derive an
analytic relation between the line flux and the apparent UV magnitude:
\begin{eqnarray}
 \log_{10} F_{\rm line} = -0.4 (m_{\rm UV} - A_{\rm UV} + AB_0) 
  + \log_{10}C_{\rm line} \cr
  - \log_{10}C_{\rm UV} + \log_{10}(1-f_{\rm esc}) - \log_{10}(1+z)\,,
\end{eqnarray}
where $F_{\rm line}$ is the line flux in cgs unit, $m_{\rm UV}$ is the
apparent UV magnitude in the AB system, $A_{\rm UV}$ is the UV dust
attenuation, $AB_0=48.6$ is the zero point of the AB system in cgs unit, 
$f_{\rm esc}$ is the LyC escape fraction, $z$ is the redshift, 
$C_{\rm line}$ is the line emissivity in Table~1, and $C_{\rm UV}$ is
the ratio of the UV luminosity-to-the SFR.  
For $>100$ Myr constant SFR and the Salpeter IMF in \S2, we obtain
the almost constant value of 
$\log_{10}(C_{\rm UV}[{\rm erg~s}^{-1}/(M_\odot~{\rm yr}^{-1})])=27.84$ 
at the rest-frame 2000 \AA\ for $Z/Z_\odot=0.02$ to 1.0. In Figure~2, we
show 3 cases of the analytic relations. The metallicity and attenuation
values are taken from the results of \cite{Shimizu2013}. From this
comparison, the readers may confirm that the galaxies are already
enriched to $Z_{\rm neb}=0.01$--$0.5Z_\odot$ and typically
$\sim0.2Z_\odot$ at these high-$z$.

In Figure~3, we show the expected peak flux density of the [O {\sc iii}]
88 $\mu$m line as a function of the apparent magnitude. The
one-dimensional velocity dispersion of the dark matter particles
composing the simulated LBGs is $42\pm4$ (or $26\pm3$) km s$^{-1}$ for
$H_{160}=27$ (28.5). This is somewhat smaller than those of optical [O
{\sc iii}] lines measured in $z\sim3$ LBGs \citep[e.g.,][]{Pettini1998}.
However it is reasonable given a lower mass of the $z\geq7$ LBGs as
\cite{Shimizu2013} expect $<10^{11}$ $M_\odot$ for $H_{160}>27$.
We expect $1.3\pm0.5$ (or $0.2\pm0.1$) mJy for $H_{160}=27$ (28.5) 
objects.\footnote{Other physical parameters of the simulated galaxies
are as follows: SFR of $20\pm7$ ($3.0\pm1.5$) $M_\odot$ yr$^{-1}$,
``nebular'' metallicity of $0.19\pm0.07$ ($0.10\pm0.05$) $Z_\odot$, 
halo mass of $(1.6\pm0.3)\times10^{11}$ ($(4.5\pm1.1)\times10^{10}$)
$M_\odot$, and stellar mass of $(2.8\pm1.1)\times10^{9}$ 
($(4.8\pm2.4)\times10^{8}$) $M_\odot$ for $H_{160}=27$ (or 28.5) AB.} 
If the readers require to estimate the strengths of other emission
lines, they can do by a scaling with the numbers in Table~1.

\begin{figure}
 \epsscale{0.7}
 \plotone{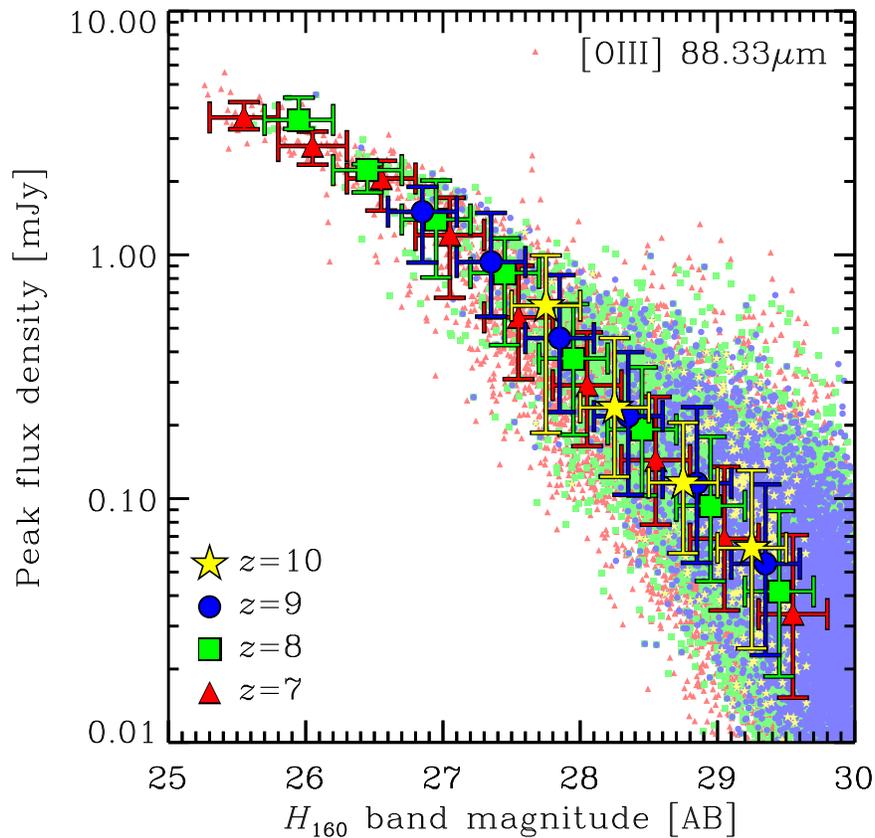}
 \caption{The expected peak flux density of the [O {\sc iii}] 88 $\mu$m
 line of Lyman break galaxies at $z\sim7$ (triangles), 8 (squares), 9
 (circles), and 10 (stars) in the simulation of \citet{Shimizu2013}. The
 large symbols with error-bars show the median values and the central
 68\% ranges.}
\end{figure}

\section{Follow-up feasibility with ALMA}

Let us consider observing the [O {\sc iii}] 88 $\mu$m line with ALMA. 
In the band 7 (275--373 GHz), we can capture the line from  
$8.1\leq z \leq11.3$ where the current highest-$z$ LBGs reside. Using
the ALMA Sensitivity Calculator for the Cycle 2, in which we have
assumed 50 antennas (full operation), dual polarization, 30 km s$^{-1}$ 
velocity resolution, declination of $+22^{\rm d}25'$ (the $z=9.6$ object
by \citealt{Zheng2012}), and `Automatic Choice' for weather
conditions\footnote{The precipitable water vapor is automatically
selected depending on the frequency by the Observing Tool for Cycle 
2 as shown in the lower panel of Figure~4. This is the
default setting and is related to the real operation of the
observatory.}, we obtain the expected sensitivities in Figure 4. We
find that the 1.3 mJy [O {\sc iii}] line from a $H_{160}=27$ object at
$8.3 < z < 9.3$ or $9.6 < z < 11.5$ can be detectable at a $>4$-$\sigma$
significance with about 1 hour integration. Since there is a
rather strong atmospheric water absorption, we cannot easily detect the
line from $9.3<z<9.6$. In fact, $H_{160}=27$ is very bright for LBGs at
$z>8$ but there are some objects found in the recent survey
\citep[e.g.,][]{Trenti2011,Oesch2013}. Gravitationally lensed objects
are also good targets. \cite{Zheng2012} reported an object with the
photometric redshift $z=9.6$. This object is as bright as $H_{160}=25.7$
apparently but should be 28.6 without magnification. According to Figure
3, we find the intrinsic and lensed line flux densities of this object
are about 0.2 and 3 mJy, respectively. Therefore, we can detect the [O
{\sc iii}] line at $>5$-$\sigma$ from this object with only 15 minutes
integration.

\begin{figure}
 \epsscale{0.7}
 \plotone{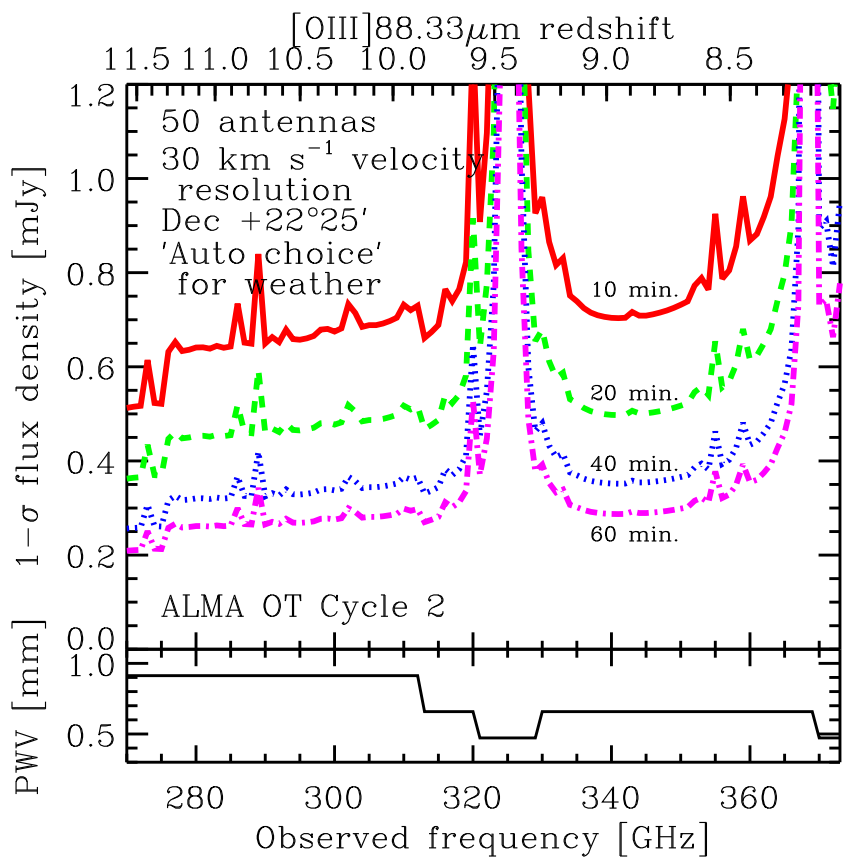}
 \caption{The expected ALMA sensitivities in the band 7 with
 a 10 (solid), 20 (dashed), 40 (dotted), and 60 (dot-dashed) minutes
 integration under the condition described in the panel. The lower panel
 shows the precipitable water vapor (PWV) values chosen by the ALMA
 Observing Tool `Automatic Choice'.}
\end{figure}

It is a caveat that we do not know exact redshift prior to the
detection. So, we have to scan a range of frequency where a possible
line exists. After ALMA Cycle 2, we can use the ``Spectral Scan'' mode
which enables us to cover about 30 GHz by 5 tunings. This corresponds to
about $\Delta z\simeq1$ for the [O {\sc iii}] line at $z\simeq9$. This
is wide enough to probe the redshift range expected by photometry. 
Another caveat is that we have only a single line even if we 
detect. Thus, we have to rely on a photometric redshift method to
conclude the detected line to be the [O {\sc iii}] 88  $\mu$m line. 
Fortunately, we can go to detect other weaker lines such as 
[O {\sc iii}] 52 $\mu$m and [N {\sc iii}] 57 $\mu$m as well as 
[O {\sc i}] 63 $\mu$m and [C {\sc ii}] 158 $\mu$m because we already
know the exact redshift and can invest much time for one integration.
The ratios of these FIR lines are very useful for diagnostics of the
chemical evolution and ionizing sources in highest-$z$ galaxies.

\acknowledgments

We thank to Tohru Nagao for useful comments on {\sc cloudy} calculations
and Erik Zackrisson for discussions. We acknowledge the financial
support of Grant-in-Aid for Young Scientists (A: 23684010; S: 
20674003; B: 24740112) by MEXT, Japan. This work is also supported by
the FIRST program SuMIRe by the Council for Science and Technology
Policy.


\begin{thebibliography}{}
\bibitem[Abel et al.(2005)]{Abel2005}
Abel, N. P., Ferland, G. J., Shaw, G., \& van Hoof, P. A. M., 
2005, ApJS, 161, 65

\bibitem[Brammer et al.(2013)]{Brammer2013}
Brammer, G. B., et al., 2013, ApJ, 765, L2

\bibitem[Brauher et al.(2008)]{Brauher2008}
Brauher, J. R., Dale, D. A., Helou, G., 2008, ApJS, 178, 280

\bibitem[Bunker et al.(2013)]{Bunker2013}
Bunker, A. J., et al., 2013, MNRAS, 430, 3314

\bibitem[Capak et al.(2013)]{Capak2013}
Capak, P. L., Faisst, A., Vieira, J. D., Tacchella, S., Carollo, M., 
\& Scoville, N. Z. 2013, ApJ, in press (arXiv:1307.4089)

\bibitem[Cormier et al.(2012)]{Cormier2012} 
Cormier, D., et al., 2012, A\&A, 548, 20

\bibitem[Ellis et al.(2013)]{Ellis2013} 
Ellis, R. S., et al., 2013, ApJ, 763, L7

\bibitem[Ferkinhoff et al.(2010)]{Ferkinhoff2010} 
Ferkinhoff, C., Hailey-Dunsheath, S., Nikola, T., Parshley, S. C.,
Stacey, G. J., Benford, D. J., \& Staguhn, J. G. 2010, ApJ, 714, L147

\bibitem[Ferland et al.(2013)]{Ferland2013} 
Ferland, G. J., et al., 2013, RMXAA, 49, 137

\bibitem[Fioc \& Rocca-Volmerange(1997)]{PEGASE} 
Fioc M., \& Rocca-Volmerange, B. 1997, A\&A, 326, 950

\bibitem[Hirashita et al.(2003)]{Hirashita2003} 
Hirashita, H., Buat, V., \& Inoue, A. K. 2003, A\&A, 410, 83

\bibitem[Inoue et al.(2000)]{Inoue2000} 
Inoue, A. K., Hirashita, H., \& Kamaya, H. 2000, PASJ, 52, 539

\bibitem[Inoue et al.(2001)]{Inoue2001a} 
Inoue, A. K., Hirashita, H., \& Kamaya, H. 2001, ApJ, 555, 613

\bibitem[Inoue(2001)]{Inoue2001b} 
Inoue, A. K. 2001, AJ, 122, 1788

\bibitem[Inoue(2011)]{Inoue2011} 
Inoue, A. K. 2011, MNRAS, 415, 2920

\bibitem[Kaneker et al.(2013)]{Kaneker2013} 
Kaneker, N., Wagg, J., Ram Chary, R., \& Carilli, C. 2013, ApJ, 771, L20

\bibitem[Kawada et al.(2011)]{Kawada2011} 
Kawada, M., et al., 2011, PASJ, 63, 903

\bibitem[Kennicutt(1998)]{Kennicutt1998}
Kennicutt, R. C., 1998, ARA\&A, 36, 189

\bibitem[Komatsu et al.(2011)]{WMAP} 
Komatsu E., et al., 2011, ApJS, 192, 18

\bibitem[Leitherer et al.(1999)]{Leitherer1999} 
Leitherer, C., et al., 1999, ApJS, 123, 3

\bibitem[Madden et al.(2012)]{Madden2012} 
Madden, S. C., et al., 2012, IAU Symposium, 284, 141

\bibitem[Madden et al.(2013)]{Madden2013} 
Madden, S. C., et al., 2013, PASP, 125, 600

\bibitem[Maiolino et al.(2005)]{Maiolino2005} 
Maiolino, R., et al., 2005, A\&A, 440, L51

\bibitem[Matsuo et al.(2009)]{Matsuo2009} 
Matsuo, H., Arai, T., Nitta, T., \& Kosaka, A. 2009, ASP Conference Series, 418, 451

\bibitem[Mizutani et al.(2002)]{Mizutani2002} 
Mizutani, M., Onaka, T., \& Shibai, H. 2002, A\&A, 382, 610

\bibitem[Nagao et al.(2011)]{Nagao2011} 
Nagao, T., Miolino, R., Marconi, A., \& Matsuhara, H. 2011, A\&A, 526, A149

\bibitem[Oesch et al.(2013)]{Oesch2013} 
Oesch, P. A., et al., 2013, arXiv:1309.2280

\bibitem[Okamoto et al.(2008)]{Okamoto2008} 
Okamoto T., Nemmen R. S., \& Bower R. G. 2008, MNRAS, 385, 161

\bibitem[Okamoto \& Frenk(2009)]{Okamoto2009} 
Okamoto T., \& Frenk C. S. 2009, MNRAS, 399, L174

\bibitem[Okamoto et al.(2010)]{Okamoto2010} 
Okamoto T., Frenk, C. S., Jenkins, A., \& Theuns, T. 2010, MNRAS, 406, 208

\bibitem[Oke(1990)]{Oke1990} 
Oke, J. B. 1990, AJ, 99, 1621

\bibitem[Ouchi et al.(2013)]{Ouchi2013} 
Ouchi, M., et al., 2013, ApJ, submitted (arXiv:1306.3572)

\bibitem[Pettini et al.(1998)]{Pettini1998}
Pettini, M., Kellogg, M., Steidel, C. C., Dickinson, M., 
Adelberger, K. L., \& Giavalisco, M. 1998, ApJ, 508, 539

\bibitem[Pirzkal et al.(2013)]{Pirzkal2013} 
Pirzkal, N., Rothberg, B., Ryan, R., Coe, D., Malhotra, S., Rhoads, J., 
\& Noeske, K. 2013, ApJ, in press (arXiv:1304.4594)

\bibitem[Schaerer(2002)]{Schaerer2002} 
Schaerer, D. 2002, A\&A, 382, 28

\bibitem[Schenker et al.(2013)]{Schenker2013} 
Schenker, M. A., et al., 2013, ApJ, 768, 196

\bibitem[Shimizu et al.(2013)]{Shimizu2013} 
Shimizu I., Inoue, A. K., Okamoto T., \& Yoshida, N. 2013, MNRAS,
submitted (arXiv:1310.0114)

\bibitem[Springel et al.(2001)]{Springel2001}
Springel, V., White, S. D. M., Tormen, G., Kauffmann, G., 2001, MNRAS, 
328, 726

\bibitem[Springel(2005)]{Gadget} 
Springel V. 2005, MNRAS, 364, 1105

\bibitem[Tremonti et al.(2004)]{Tremonti2004}
Tremonti, C. A., et al., 2004, ApJ, 613, 989

\bibitem[Trenti et al.(2011)]{Trenti2011}
Trenti, M., et al., 2011, ApJ, 727, L39

\bibitem[Treu et al.(2013)]{Treu2013} 
Treu, T., Schmidt, K. B., Trenti, M., Bradley, L. D., \& Stiavelli, M. 
2013, ApJ, accepted (arXiv:1308.5985)

\bibitem[van den Bergh(2000)]{vdBergh2000}
van den Bergh, S., 2000, The Galaxies of the Local Group, 
Cambridge University Press

\bibitem[Walter et al.(2012)]{Walter2012} 
Walter, F., et al., 2012, ApJ, 752, 93

\bibitem[Ward et al.(1975)]{Ward1975} 
Ward, D. B., Dennison, B., Gull, G., \& Harwit, M. 1975, ApJ, 202, L31

\bibitem[Zheng et al.(2012)]{Zheng2012}
Zheng, W., et al., 2012, Nature, 489, 406
\end{thebibliography}
\end{document}